# Low temperature 2D GaN growth on Si (111) 7x7 assisted by hypethermal nitrogen ions

*Jaroslav Maniš[(1,2)]\*, Jindřich Mach[(1,2)], Miroslav Bartošík [(1,2,3)], Tomáš Šamořil[(2)], Michal Horak[(2)], Vojtěch Čalkovský[(2)], David Nezval[(2)], Lukáš Kachtik[(1)], Martin Konečny[(1,2)], and Tomáš Šikola[(1,2)]*

[1] CEITEC BUT, Brno University of Technology, Technická 3058/10, 616 00 Brno, Czech Republic

[2] Institute of Physical Engineering, Brno University of Technology, Technická 2, 616 69 Brno, Czech Republic

[3] Department of Physics and Materials Engineering, Faculty of Technology, Tomas Bata University in Zlín, 760 01, Czech Republic

\* Corresponding author: mach@fme.vutbr.cz

KEYWORDS: Ga, GaN, 2D nanostructures, low-temperature droplet epitaxy (LTDE), nitrogen ions, ion-atom beam, TEM, SEM, Auger

ABSTRACT:

*As the characteristic dimensions of modern top-down devices are getting smaller, such devices reach their operational limits given by quantum mechanics. Thus, two-dimensional (2D) structures appear as one of the best solutions to meet the ultimate challenges of modern optoelectronic and spintronic applications. The representative of III-V semiconductors, gallium nitride (GaN), is a great candidate for UV and high-power applications at a nanoscale level. We propose a new way of fabrication of 2D GaN on the Si(111) 7×7 surface using post-nitridation of Ga droplets by hyperthermal (E < 50 eV) nitrogen ions at low substrate temperatures (T < 220°C). Both deposition of Ga droplets and their post-nitridation is carried out using a special atom-ion beam source developed in our group. This low-temperature droplet epitaxy (LTDE) approach provides well-defined ultra-high vacuum growth conditions during the whole fabrication process resulting in high purity GaN nanostructures. A sharp*

*interface between the GaN nanostructures and the silicon substrate together with a correct elemental composition of nanostructures was confirmed by TEM. In addition, SEM, X-ray photoelectron spectroscopy (XPS), AFM and Auger microanalysis showed unique characteristics of the fabricated GaN nanostructures.*

1. **INTRODUCTION**

Motivated by the discovery of graphene, scientists have made a great effort to extend their knowledge and fabricate other two-dimensional (2D) materials. Among others, 2D forms of III-V semiconductor nanostructures, such as gallium nitride (GaN) layers represent a suitable material for optical applications. The GaN based ultra-violet laser diodes are used to water cleaning [1], gas detoxification [2],[3],[4] or as a light source for optical disc recording technologies. The very high breakdown voltage and high electron mobility also make it an ideal material for high-performance, high-frequency devices (for example, high-speed wireless data transmission, high-voltage switching).

In addition, 2D semiconductors are one of the possible candidates in overcoming the short-channel effects (mitigating heat dissipation) in FET transistors [5], [6], which could improve the efficiency of these devices greatly. Recently, an excellent storage capacity has been reported when 2D GaN was used as an anode in Li-ion and Na-ion batteries [7]. The ferromagnetic properties of 2D GaN have been studied from the perspective of doping by copper [8] and other elements [9], showing large spin polarization and high Curie temperature. The role of vacancies in electrical properties [10] as well as the impact of buckling of a 2D GaN structure on carrier mobility [11] have been already investigated with promising results, thus paving the way for implementing 2D GaN nanostructures in spintronic nanodevices.

The first successful attempt to fabricate two-dimensional gallium nitride was achieved by a migration enhanced epitaxial growth (MEEG) method [12]. 2D GaN was grown sandwiched in between graphene and a passivated silicon substrate. The authors suggested that graphene plays the crucial role in stabilizing the direct 5eV GaN bandgap. The MEEG technique also allows us to control the growth rate of 2D GaN structure resulting in the precise control of the number of formed layers. It has been shown that the number of layers has a crucial role in controlling the crystallographic structure which further influences the band gap of the GaN structure [13]. Even though MEEG method enables to grow 2D GaN, the sandwiched structure limits further applications since it hinders the possibility of transferring the fabricated nanostructure to other materials.

Recently, researchers have developed the method for fabrication of micrometer-sized 2D GaN structures of ultimate thickness [14]. Heat decomposition of a nitrogen precursor and subsequent reaction of nitrogen with liquefied gallium spread over a tungsten substrate leads to the formation of a uniform 2D GaN layer. Reported 2D GaN exhibits promising electrical and optical properties along with the possibility of transferring the structure to other materials as demonstrated by a fabricated functional device.

However, all referred 2D nanostructures were fabricated by MOCVD methods which has several disadvantages. Firstly, high temperatures and hazardous precursors which are inflammable and poisonous have been generally used. Secondly, the impurities can be introduced into the formed nanostructure during the growing process due to a higher base pressure which usually reaches tens of Pascals. Since impurities generally affect the electronic and optical properties significantly (e.g. as dopants), it is highly demanding to find fabrication techniques enabling to reduce additional elements significantly. In addition, the MOCVD technique is not suitable for real-time *in situ* surface analysis of grown nanostructures.

The droplet epitaxy (DE) is an alternative fabrication method of nanostructures which can be carried out in standard MBE vacuum chambers [15]. Even though DE was originally developed for the fabrication of GaAs nanodots [16], in the paper we show how to adopt this method for growth of GaN nanostructures. The DE method brings the advantage of a precise control of the ambient environment since it introduces only the elements of interest and allows us to conduct the fabrication process at ultra-high vacuum (UHV) conditions. On top of that, the precise control of the deposition rate enables to achieve the required nanostructure size. We propose to adopt DE for low-temperature synthesis of GaN by using a special ion-atom beam source developed in our group [17][18]. In this way, a low-temperature droplet epitaxy (LTDE) method under the well-defined conditions can be utilized for fabrication of 2D GaN nanostructures with a well defined crystal structure. Such a low-temperature growth opens the way for fabrication of photoactive 2D GaN nanostructures in the vicinity of the already pre-fabricated metallic plasmonic antennas, thus enabling optimization of plasmon-enhanced photoluminescence effects and other plasmon-based phenomena (e.g. strong coupling).

2. **RESULTS AND DISCUSSION**

The fabrication of 2D GaN nanostructures using the droplet epitaxy method is a two-step process. Firstly, the deposition of metallic gallium results in a formation of gallium droplets uniformly spread over the Si(111) 7×7 surface. Secondly, subsequent nitridation leads to a transformation of Ga droplets into GaN nanocrystals. This process is described in detail in [19] and also in our paper [20]. The size and density of Ga droplets on the Si(111) 7×7 surface can be controlled both by the surface temperature and deposition time [21]. Ga deposition on Si(111) 7×7 was carried out for 50 min (resulted in a coverage of 5 ML) at 330°C. The density of gallium droplets spread over the Si(111) 7×7 surface was approximately $3 \times 10^8$ droplets per $cm^2$. For these deposition conditions a droplet diameter of 200 nm and height of 70 nm have been observed by AFM. Afterwards, the substrate temperature was decreased to 210°C and gallium droplets were exposed to a hyperthermal nitrogen ion beam ($E = 50$ eV) for 120 min to accomplish post-nitridation. The temperature during the post-nitridation process was the critical parameter of the experiment since the growth of 2D GaN was observed only in a narrow low temperature window (190°C-210°C).

Fig. 1a) shows a SEM image of a triangle-like 2D GaN nanostructure. This nanostructure grew in the direct contact with a circular object being the initial gallium droplet. The bright feature between the Ga droplet and the GaN triangle is most likely bulky GaN. The shape of 2D GaN nanostructures in the form of an equilateral triangle indicates epitaxial growth of GaN being governed by the Si(111) 7×7 surface reconstruction possessing the six-fold hexagonal symmetry. This is a completely different result with respect to the low-temperature GaN crystal growth on the native $SiO_2$ surface performed in our group [20], where Ga droplets were transformed into 3D GaN nanocrystals. The growth process of GaN structures was investigated by changing the time of nitridation as illustrated in Fig. 1 b). Initially, after the gallium deposition, the silicon surface is covered by liquefied gallium droplets. Once the nitridation is initiated, incoming nitrogen ions start to react with metallic gallium which leads to the formation of a triangle-like GaN structure in the up-stream direction with respect to the nitrogen flux surface projection (indicated by the arrow). During this process, the decreasing gallium droplet is asymmetrically dewetting from the silicon surface towards the growing GaN structure. As gallium is gradually consumed in the GaN synthesis process, the triangle-like shaped structure is becoming thinner and more prominent. It is obvious that Ga atoms must diffuse across the GaN surface to feed the reaction with nitrogen at the GaN border. It is obvious that the size of the GaN structure is limited by the amount of the source material, i.e. by the size of the initial gallium droplet. The resultant GaN nanostructures are as thin as 5 nm and uniformly flat as was observed by the AFM. Thus, we call them 2D nanostructures.

Once the gallium is fully consumed, the original gallium droplet vanishes, leaving the deteriorated silicon surface behind. This phenomenon, referred as the meltback etching and caused by a chemical reaction of gallium with silicon, has been also reported by [22] and other groups for MOCVD growth of GaN crystals. This etching does not occur in the area of 2D GaN growth because it is most likely outperformed by the Ga reaction with nitrogen species being directly delivered via a hyperthermal nitrogen ion beam to the reaction zone at the GaN border. On the other hand, the left-hand side of the Ga droplet is screened off from the nitrogen flux and thus the Ga etching to the GaN reaction prevails. This explanation is supported by the finding in [22] that addition of $N_2$ to the carrier gas blocks meltback etching.

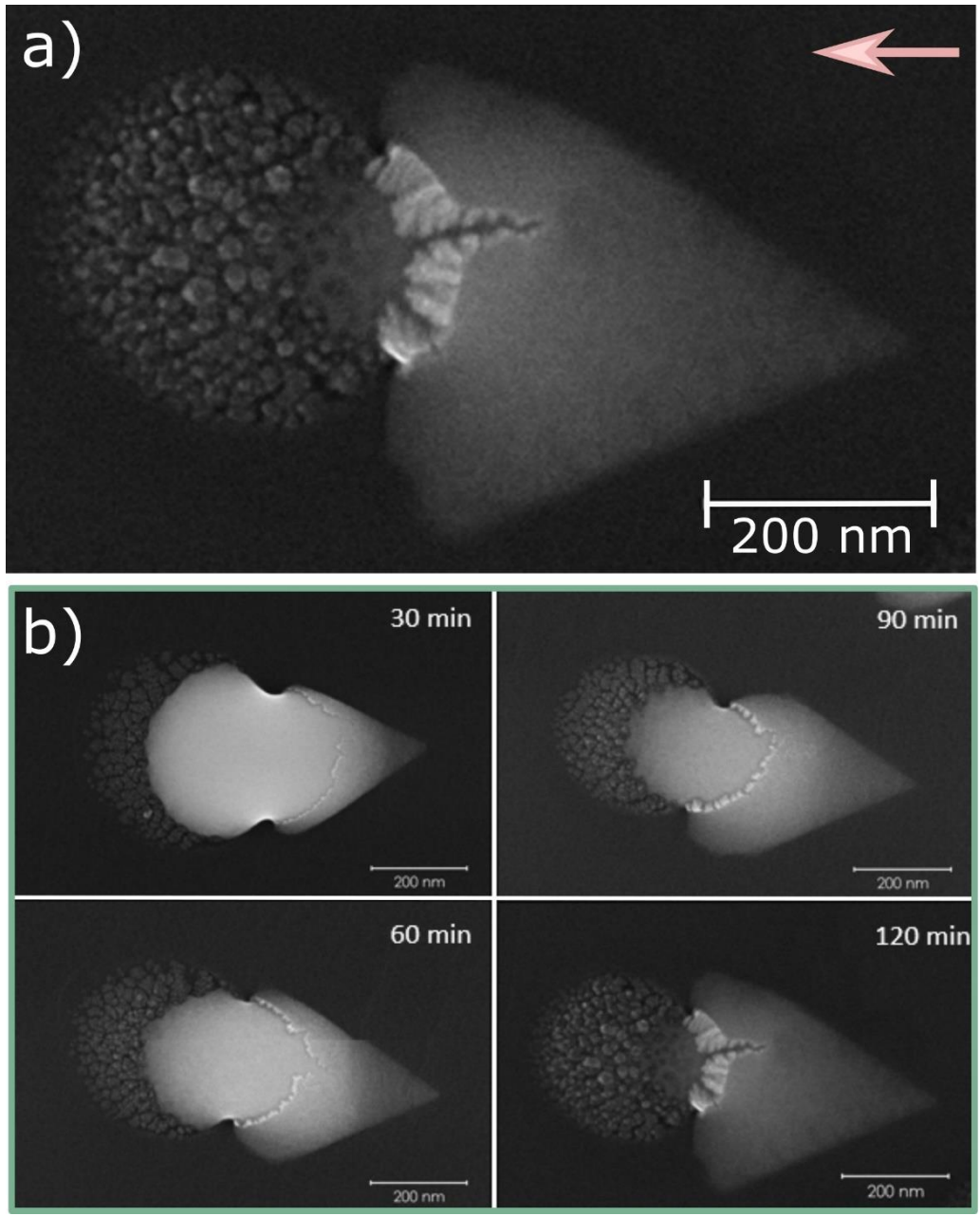

**Fig. 1.** a) SEM image of a 2D GaN triangle-like nanostructure. b) Time evolution of this nanostructure upon nitridation. Increasing the time of nitridation leads to the formation of a thin triangle-like structure. Once the original gallium droplet is fully consumed, growth is finished.

The effect of post-nitridation was studied by XPS. Fig. 2a) shows a Ga $2p_{3/2}$ peak measured after 60 min of post-nitridation. The fitting of the Ga $2p_{3/2}$ peak revealed the presence of two components. The first component at 1117.8 eV corresponds to the chemical bonding between Ga atoms in a metallic form (Ga-Ga bonds). The second component at 1119.3 eV represents the chemical bonding between Ga and N in GaN crystals (Ga-N bonds). The fitting parameters of the Ga $2p_{3/2}$ peak were reported in [20]. The inserted AFM image of a Ga droplet clearly shows that this droplet is not fully transformed into 2D GaN. Fig. 2b) shows the Ga $2p_{3/2}$ peak after 120 min of post-nitridation. Both peak components – Ga-Ga and Ga-N – are still present in the spectra. However, the Ga-N component is much more prominent than the Ga-Ga one which suggests that most of the metallic gallium from Ga droplets has been transformed into the GaN compound, which is also supported by the inserted AFM image of an analyzed structure. The fragment of Ga-Ga bonding most likely comes from the area of the original droplet or from the bright feature between the original droplet and the triangle-like structure which is most likely bulky GaN.

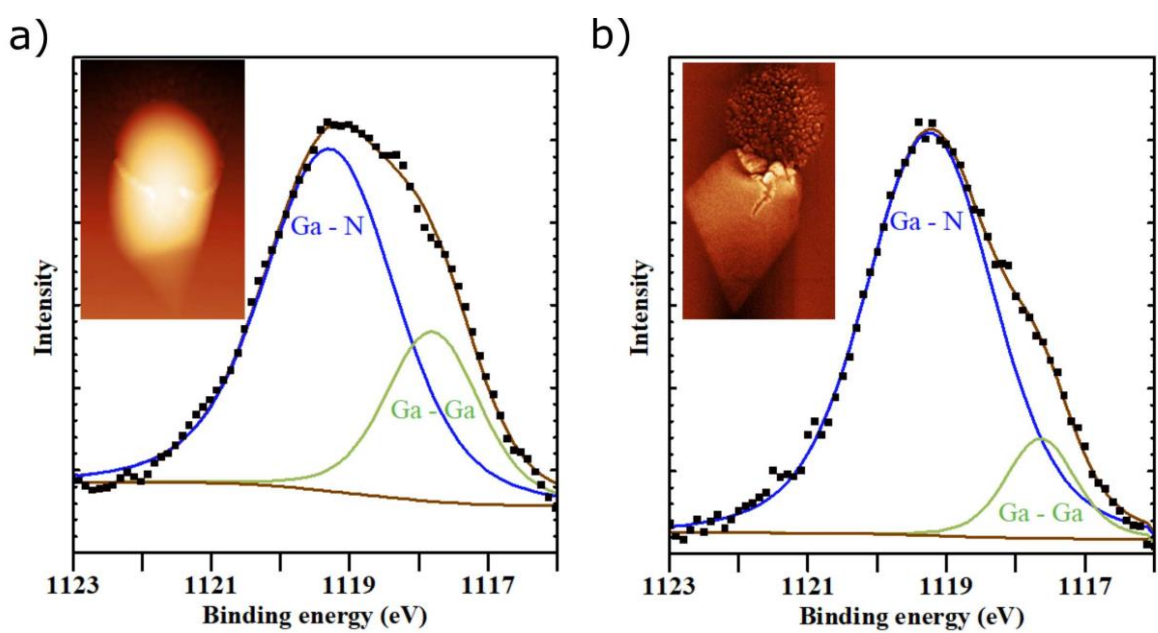

**Fig. 2.** Fitted XPS Ga $_{2p3/2}$ peak taken during the post-nitridation after a) 60 min and b) 120 min. The Ga-Ga component is at 1117.8 eV (green curve) and Ga-N component at 1119.3 eV (blue curve).

To elucidate the elemental components distribution in the individual nanostructures, Auger nanoanalysis by NanoSAM was performed. There are three different selected regions **A**, **B** and **C** in the SEM image of the nanostructure in Fig. 3a). The region **A** corresponds to the original Si(111) surface far away from the droplet and nanostructure, the region **B** to the triangle-like nanostructure, and the region **C** represents the peripheral area of the original gallium droplet. Fig. 3b) shows the signal of Auger electrons in the range of kinetic energy corresponding to Ga atoms. Gallium is present in the triangle-like shaped nanostructure (**B**), as well as in the region of the original droplet (**C**). Gallium atoms are not present in the region **A**. Fig. 3c) shows the signal of Auger electrons in the energy range corresponding to N atoms. The presence of nitrogen in the region **A** indicates that nitrogen ions react with the pristine silicon surface and form a thin layer of silicon nitride. Nitrogen is also present in the triangle-like GaN structure (region **B**). However, there is no evidence of nitrogen in the area of original gallium droplet (region **C**). Such results lead us to the conclusion that nitrogen ions do not penetrate into the Ga droplet and thus do not form Ga-N bonds here. As proposed by Gerlach et al. [19], nitrogen ions rather diffuse along the liquid-vapor interface (Ga droplet surface) to peripheral regions of the droplet – substrate interface (triple point, i.e. the solid-liquid-vapor interface). At this interface, the nucleation of GaN is initiated which eventually results in the formation of a 2D GaN layer.

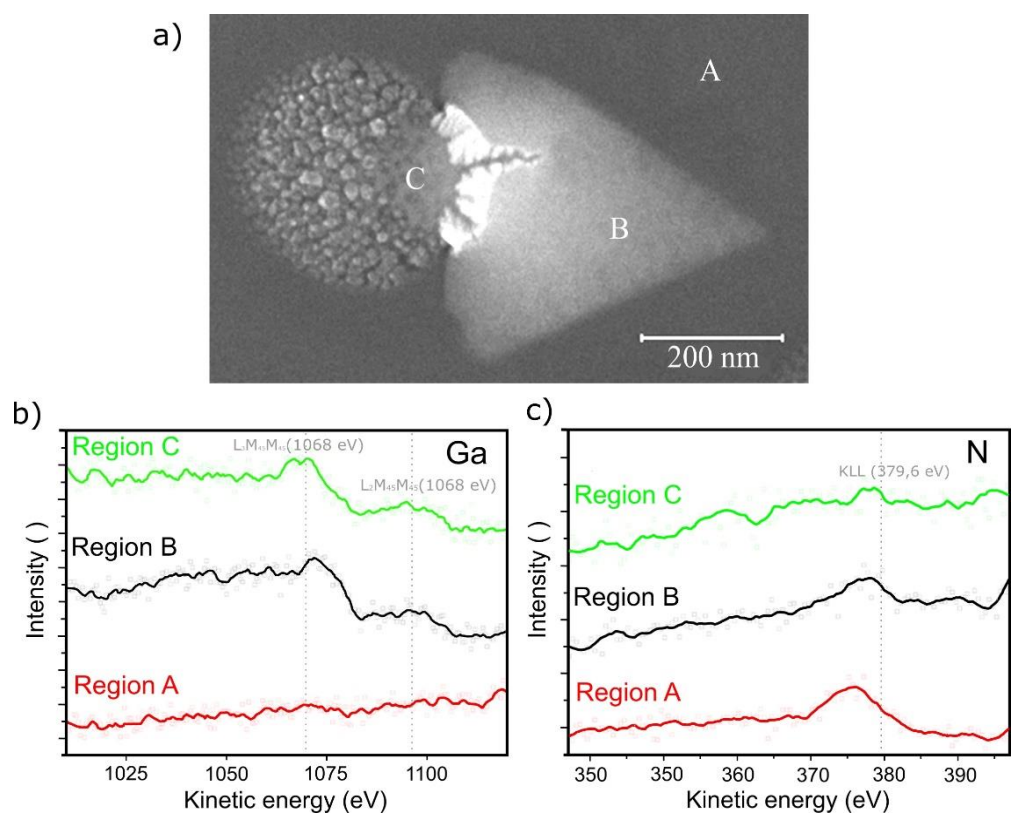

**Fig. 3.** Elemental analysis of a triangle-like GaN nanostructure carried out by NanoSAM. A) SEM image of the studied structure with the selected regions A, B and C. b) Gallium and c) nitrogen abundance in the selected regions.

STEM image of a cross-sectional lamella (Fig. 4a) reveals sharp interfaces between the Si(111) substrate and the amorphous gallium droplet, and between this substrate and the crystalline hexagonal GaN. Fig.

4b) shows zoom-in detail of the 2D GaN-Si(111) interface. The observed interlayer distance 3.15±0.09 Å in the 2D GaN structure is higher than a value of 2.59 Å found experimentally in wurtzite GaN crystals [17], [23]. The calculated interlayer distance for layered 2D GaN varied from 2.22 Å [24] to 3.10 Å [25], [26], depending on the type of numerical simulation used. Unlike van der Waals structures such as graphite and others, 2D GaN transforms into different crystalline forms due to the forming of vertical bonds between Ga and N between different layers. Thus, more than three different stable types of 2D GaN stacking was predicted [25], and, hence, a variety of different interlayer distances can be observed. The measured lattice parameter $a$ was 3.29±0.13 Å which is slightly higher than in wurtzite GaN (3.18 Å) [27], [23], [28]. However, it is very close to calculated values for 2D GaN (3.28 Å) [29].

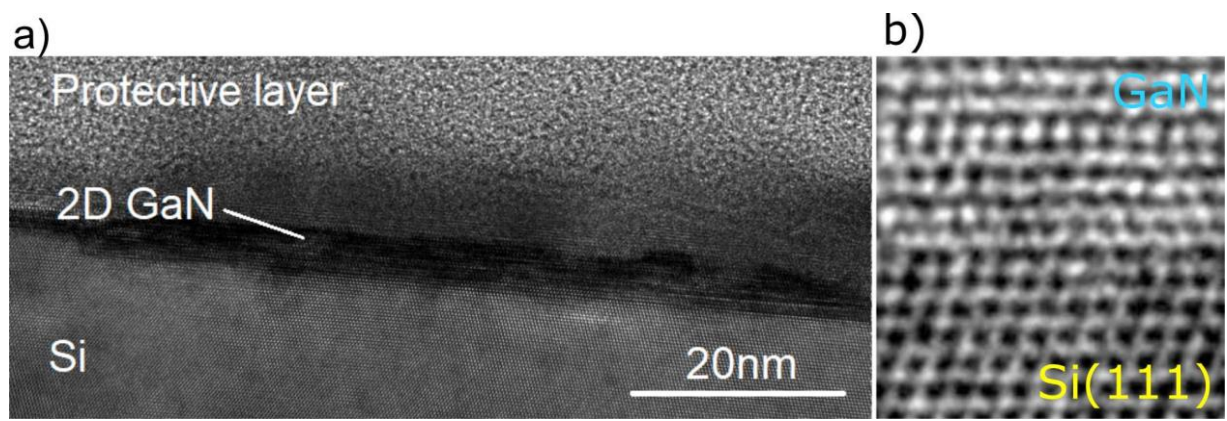

**Fig. 4.** a) STEM image of the cross-sectional lamella of an amorphous gallium droplet and crystalline GaN on the Si(111) 7×7 substrate. b) Zoom-in STEM image of the 2D GaN -Si(111) 7×7 - interface.

Fig. 5 shows an EDX measurement of gallium and nitrogen elemental abundance in the 2D GaN structure. While gallium is detected in the whole structure, i.e. in the original gallium droplet and in the formed 2D GaN layered nanostructure, nitrogen is present exclusively in the region of 2D GaN. Such a result corresponds to the Auger microanalysis results discussed above and supports the conclusion that nitrogen atoms are not present in the gallium droplet.

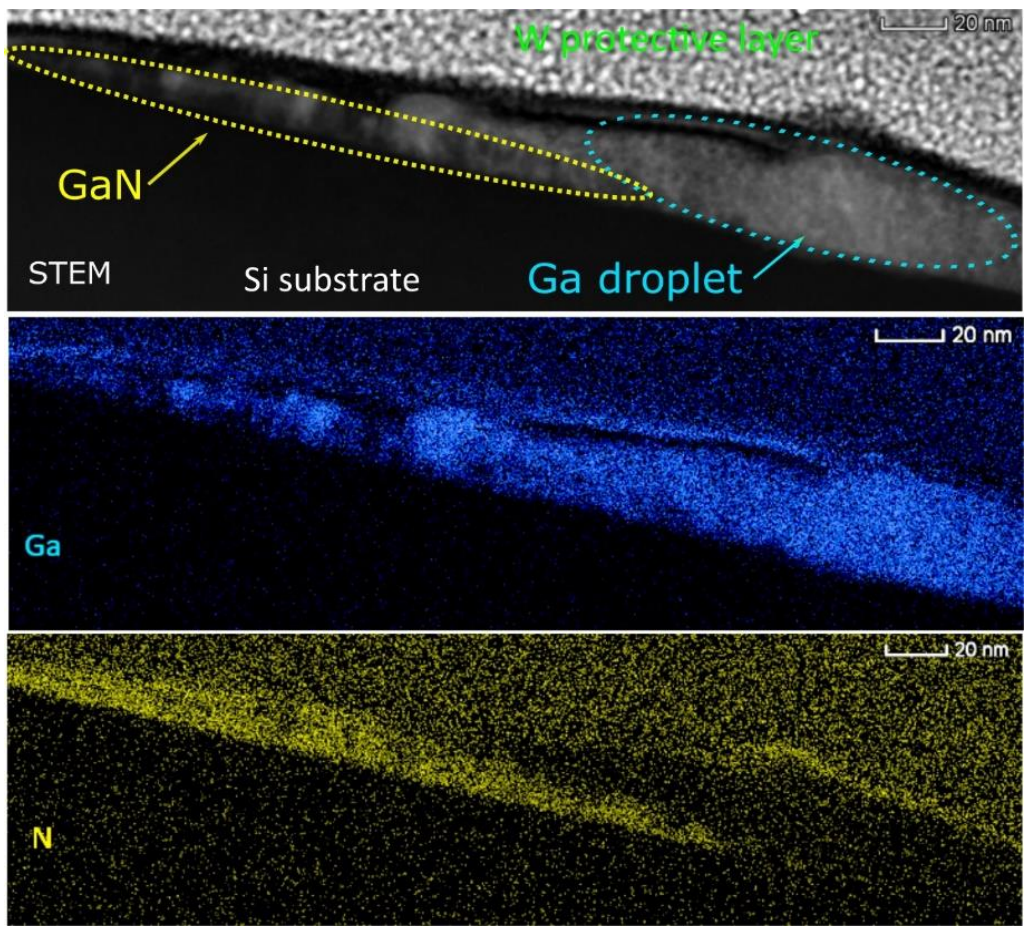

**Fig. 5.** EDX maps of the cross-sectional lamella of an amorphous gallium droplet and crystalline GaN on the Si(111) 7x7 substrate.. While gallium is present in the crystalline structure (area encircled by the yellow loop) as well as in the amorphous droplet (blue loop), nitrogen is present exclusively in the crystalline 2D GaN nanostructure.

3. **CONCLUSION**

The 2D GaN nanostructures were fabricated using a low temperature droplet epitaxy with the assistance of hyperthermal nitrogen ions. The Auger nanoanalysis and EDX measurement have indicated nitrogen particles do not penetrate into the Ga droplets, but diffuse along their surface towards the liquid-solid interface at the periphery of the droplet. Here, the nuclei of GaN are created which consequently leads to the formation of the triangle-like 2D GaN nanostructures. The nanostructures grow at their borders via a reaction between Ga atoms delivered here from the original droplets and incoming nitrogen species. Once the gallium is fully consumed, the original gallium droplet vanishes, leaving the deteriorated silicon surface behind. The STEM imaging of a cross-sectional lamella revealed an interlayer spacing distance of 3.16 Å which is higher than in the case of well-known wurtzite GaN by 22 %, but in agreement with the presented values calculated for 2D GaN structures. A lattice parameter of 3.28 Å found by the same method is in the very good agreement with the calculated values. The 2D GaN nanostructures are ready for the exploration of their optical and electrical properties. This will become the topic of our next experiments.

4. **EXPERIMENT SECTION**

2D GaN nanostructures were fabricated by a hybrid method alternating deposition of Ga atoms and ultra-low energy (hyperthermal) nitrogen ions ($E = 50$ eV) at low substrate temperature. An n-doped Si (111) single crystal (resistivity ρ = 0.01 – 0.02 Ω.cm) with 0.2° miscut was used as a substrate. After insertion into an UHV chamber, the silicon substrate was annealed at 700°C (measured by a pyrometer) for at least 12 hours by direct current heating. Afterwards, a flashing routine, i.e. a periodic rapid increase and decrease of temperature up to 1250°C and down to 800 °C, respectively, followed by a slow temperature decrease to room temperature (RT) was carried out. Simultaneously, the pressure was kept during the whole procedure below $1 \times 10^{-6}$ Pa. The flashing routine provides two benefits – complete decomposition of a native silicon oxide layer and formation of the 7×7 surface Si reconstruction. The substrate prepared by this procedure exhibited high surface cleanliness confirmed by XPS, and a uniform surface with large terraces checked by LEED.

Gallium atoms were deposited by an effusion cell in a UHV chamber with a base pressure of $5 \times 10^{-8}$ Pa using a flux rate of $7 \times 10^{12}$ atoms.cm$^{-2}$.s$^{-1}$ which resulted in a substrate coverage of $4 \times 10^{16}$ atoms.cm$^{-2}$ (5 monolayers).

Nitridation of gallium droplets was then carried out by hyperthermal nitrogen ions ($E = 50$ eV) using the ion-atom beam nitrogen source (ion source mode) at a nitrogen partial pressure of $5.5 \times 10^{-5}$ Pa. During nitridation, the flux of nitrogen ions (N$^+$ and N$^+_2$) provided a current density of 1000 nA/cm$^2$ at a beam incidence angle of 55 °.

The morphology of 2D GaN nanostructures was studied ex-situ by scanning electron microscopy (SEM - Verios 360L system, Thermo Fisher Scientific) and atomic force microscopy (AFM – ICON, Bruker). The structural microanalysis was conducted in a complex vacuum system dedicated for scanning Auger microscopy (NanoSAM equipped with an UHV Gemini electron column). A TEM lamella for investigation of the crystallographic structure was prepared using a focused Ga ion beam system (FIB) in a SEM system (Lyra, Tescan) and the observation of crystallography as well as EDS study were performed in a TEM/STEM system (Titan, Thermo Fisher Scientific). XPS was carried out using an experimental setup consisting of X-ray source DAR400 and hemispherical electrostatic analyzer (both Omicron).

**ACKNOWLEDGEMENT**

We acknowledge the support by the Czech Science Foundation (Grant No.*20-28573S*), European Commission (H2020-Twininning project No. 810626 – SINNCE, M-ERA NET HYSUCAP/TACR-TH71020004),*BUT* – specific research No.*FSI-S-20-648*5, and Ministry of Education, Youth and Sports of the Czech Republic (CzechNanoLab Research Infrastructure - LM2018110).